\documentclass[twoside,12pt]{article}
\usepackage{epsfig}

\def\Journal#1#2#3#4{{#1} {#2} (#4) #3 }

\def\NPA{{\em Nucl. Phys.} A}

\def\PLB{{\em Phys. Lett.} B}

\def\PREP{\em Phys. Rep.}

\def\PRC{{\em Phys. Rev.} C}

\def\EPJ{{\em Euro. Phys. J.} A}
\def\ANNP{\em Ann. Phys. (N.Y.)}
\def\AnRev{\em Ann. Rev. Phys. Sci. (N.Y.)}
\def\JETP{\em JETP}
\def\AOP{\em Adv. Nucl. Phys. (N.Y.)}
\def\JPG{{\em J. Phys.} G}

\newcommand{\be}{\begin{equation}}
\newcommand{\ee}{\end{equation}}
\newcommand{\bea}{\begin{eqnarray}}
\newcommand{\eea}{\end{eqnarray}}

\begin{document}

\title{Probing Isospin Dynamics in Halo Nuclei
}

\author{H.\ Lenske, F.\ Hofmann and C.M.\ Keil
\\
Institut f\"ur Theoretische Physik,\\ Universit\"at Giessen,
D-35392 Giessen, Germany }
\date{}

\maketitle
\begin{abstract} Nuclear many-body theory is used to study nuclear
matter and finite nuclei at extreme isospin. In-medium
interactions in asymmetric nuclear matter are obtained from
(Dirac-) Brueckner theory. Neutron skin formation in Ni and Sn
isotopes is investigated by relativistic mean-field calculations
in DDRH theory with density dependent meson-nucleon vertices.
Applications to light nuclei are discussed with special emphasis
on pairing and core polarization in weakly bound nuclei.
Approaches accounting for continuum coupling in dripline pairing
and core polarization are presented. Calculations for the halo
nuclei $^8$B, $^{11}$Be and $^{19}$C show that shell structures
are dissolving when the driplines are approached. Relativistic
breakup data are well described by eikonal calculations.
\end{abstract}

\section{Introduction \label{intro}}

Tremendous progress has been made over the last decade in
extending our knowledge into hitherto unknown regions of the
nuclear chart. The experimental achievements \cite{ov} were
accompanied by complementary developments in nuclear theory.
Nuclear structure physics has now access to nuclei with a large
variety of proton-to-neutron rations, hence allowing to study
nuclear forces at extreme isospin. An equally important new aspect
is the strong reduction in separation energies close to the
driplines. Under such conditions the conventional binding
mechanisms known from $\beta$ stable nuclei are likely to cease to
be valid. There are experimental indications - which are strongly
supported by theory - that single particle shell structures become
dissolved in the vicinity of proton and neutron driplines. If
confirmed this observation will have important consequences
because it means that the seemingly well-established prevalence of
static mean-field dynamics is being replaced by dynamical
processes from nucleon-nucleon (NN) interactions. In a very clean
way this type of dynamics is observed in halo nuclei while in
medium and heavy nuclei the dominance of the nuclear mean-field
seems to survive. But with increasing neutron excess the character
of the mean-field is dramatically changed by the strong
enhancement of isovector interactions. For very neutron-rich
nuclei theory predicts the formation of rather thick layers of
almost pure neutron matter representing a genuine state of
de-mixed proton and neutron fluids.

Descriptions of nuclear interactions and nuclei far off stability
necessarily afford strong extrapolations. Since data are still
scarce the theoretical calculations rely on nuclear models, to be
justified (or falsified) by experiment. An approach attempting a
self-contained description of stable and unstable nuclei by
nuclear many-body theory is presented in the following sections.

\section{Interactions in Strongly Asymmetric Matter
\label{interact}}

An appropriate approach to in-medium nuclear interactions is
Brueckner theory. Especially the relativistic effects accounted
for with Dirac-Brueckner (DB) calculations have proven to describe
nuclear matter properties rather satisfactorily. In
ref.\cite{dejong} DB Hartree-Fock (DBHF) theory was used to
investigate interactions in infinite matter over a large range of
asymmetries. The Groningen free-space NN-potential was applied.
Effective in-medium meson-nucleon coupling constants were
extracted for the isoscalar $\sigma$ and $\omega$ and the
isovector $\rho$ and $\delta$ mesons, respectively \cite{dejong}.
In all meson channels a pronounced dependence on the isoscalar
bulk density is found while the dependence on the asymmetry is
close to negligible. Hence, to a very good approximation in-medium
strong interactions remain intrinsically independent of isospin
thus conserving a fundamental symmetry. Using the density
dependent relativistic hadron (DDRH) field theory \cite{DDRH} the
DB coupling constants have been applied to finite nuclei
\cite{HKL}, hypernuclei \cite{hyper} and neutron stars \cite{ns}.
DDRH theory is a microscopic formulation of relativistic
mean-field theory \cite{rmf} but describing in-medium effects by
density dependent coupling constants rather than by non-linear
meson self-interactions. The DDRH calculations describe binding
energies, separation energies and other ground state properties of
stable and unstable nuclei rather accurately on a level of a few
percent \cite{HKL} showing that the gross properties of nuclear
matter and finite nuclei are described very satisfactorily with
microscopic interactions.

\begin{figure}[t]
\begin{center}
\begin{minipage}[t]{8 cm}
\epsfig{file=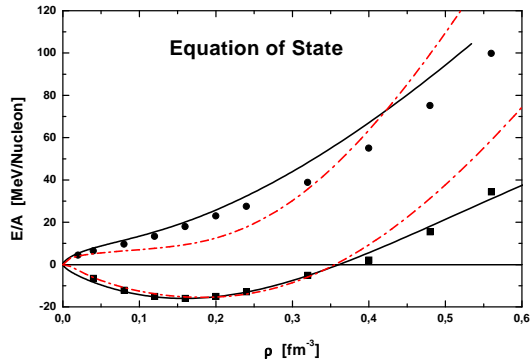,width=8cm}
\end{minipage}
\begin{minipage}[t]{16.5 cm}
\caption{Equation of state for symmetric nuclear matter and
neutron matter obtained with the D3Y interaction (full lines), the
relativistic DDRH interaction \protect\cite{HKL} (dashed-dotted
lines) and from variational calculations \protect\cite{pa} (full
squares and circles). \label{fig1}}
\end{minipage}
\end{center}
\end{figure}

A clear indication for the incompleteness of a ladder-type
interaction model is found in a re-analysis of the DDRH results
with a mass-formula showing that especially surface energy
contributions are underestimated \cite{HKL}. An important class of
interactions not accounted for by DB theory are contributions from
ring-diagrams which will become important at low densities. Also
missing are three-body interactions \cite{pa}. In order to account
for the missing contributions we follow the approach of \cite{hl}
and introduce a semi-microscopic effective interaction by
multiplying the Brueckner results by an additional density
dependent vertex functional where parameters are fixed in
symmetric matter. Results for the equation of state of symmetric
and pure neutron matter obtained with DDRH theory are displayed in
Fig.\ref{fig1} and compared to the non-relativistic density
dependent D3Y parameterization \cite{hl}.

\begin{figure}[t]
\begin{center}
\begin{minipage}[t]{8 cm}
\epsfig{file=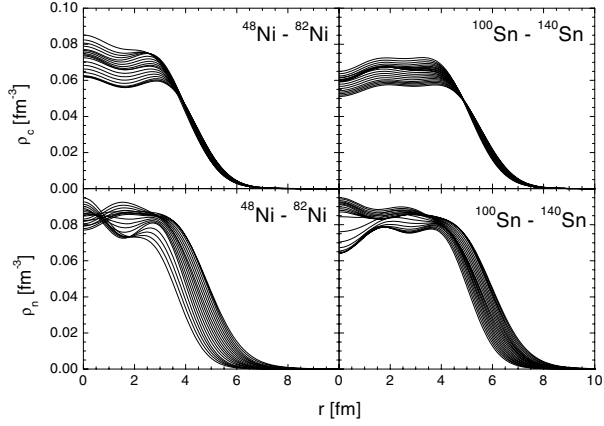,width=8cm}
\end{minipage}
\begin{minipage}[t]{16.5 cm}
\caption{Charge densities (upper panel) and neutron ground state
density distributions (lower panel) for Ni- and Sn-isotopes
\protect\cite{HKL}. The formation of neutron skins in the
neutron-rich nuclides is clearly visible. \label{fig1a}}
\end{minipage}
\end{center}
\end{figure}

An interesting effect, predicted by many mean-field models, is the
appearance of neutron- and proton- skins when the driplines are
approached. In Fig.\ref{fig1a} DDRH proton and neutron density
distributions for the isotopic chains of $^{48-82}$Ni and
$^{100-140}$Sn nuclides are displayed \cite{HKL}. At the proton
dripline ($^{48}$Ni, $^{100}$Sn ) proton skins are found being
mainly caused by the Coulomb repulsion. With increasing neutron
excess neutron skins starts to develop which, for example in the
Ni-case, reach a thickness of about $\Delta=0.8$~fm at the neutron
dripline, obtained here for $^{92}$Ni. In skin nuclei a core of
normal composition is coated by a layer of almost pure neutron
matter. Investigation of this very particular state of nuclear
matter by transfer reactions at REX-ISOLDE energies is discussed
in \cite{transfer}.

In most of the following calculations for dripline nuclei
non-relativistic HFB and RPA (or QRPA) theory will be used. The
main reason for doing so is that at present the extension of DDRH
theory (as is also true for phenomenological RMF models) to a full
dynamical description is still in progress. Relativistic
interactions are well tested at the level of HF and HFB theory,
i.e. in the limit of a static ground state calculations in the
{\it no-sea} approximation. But only few is known about extensions
to dynamical processes, which in a relativistic approach should
take into account also the coupling to vacuum modes.

The non-relativistic D3Y interaction is based on a
parameterization of the G-Matrix in terms of boson exchange
interactions, including $\pi$, $\sigma$, $\omega$ and $\rho$
exchange. Because of the known inability of non-relativistic
Brueckner theory to reproduce adequately the empirical nuclear
matter EoS an additional density dependent re-scaling was
introduced in the {\it particle-hole (ph)} channel, enforcing
agreement with the variational EoS of ref.\cite{pa} of for
symmetric nuclear matter. Using Landau-Migdal theory
\cite{physrep} the appropriate (density dependent) $ph$
interactions in the various spin and isospin channels are derived
self-consistently.

\section{Pairing at the Particle Threshold}

The existence of $^{11}$Li as a particle-stable system relies
almost completely on the mutual interactions among the last two
valence neutrons. Their low separation energy, S$_{2n}$=320~keV,
points to the importance of the mixing with unbound states. For a
detailed description of the valence wave function the conventional
BCS and HFB methods using representations in terms of mean-field
wave functions are not suitable. Theoretically, continuum effects
are properly described by solving the Gorkov-equations \cite{go}
\begin{equation} \label{gorkov}
(h-e_{+})\Phi_{+} - \Delta \Phi_{-} = 0 \quad , \quad
(h-e_{-})\Phi_{-} + \Delta \Phi_{+} = 0
\end{equation}
as coupled equations for the hole and particle components
\cite{le} which are denoted by $\Phi_\pm$ and are coupled by the
pairing field $\Delta$. The single particle energies are
$e_\pm=\lambda \pm E$ where $\lambda$ and E ($\geq 0$) are the
chemical potential and the quasiparticle energy, respectively.
Note, that $e_\pm$ will in general {\it not} coincide with the
eigenvalues of the mean-field Hamiltonian $h$. In this sense, the
states $\Phi_\pm$ are off the (mean-field) energy shell and
deviate from usual quasiparticle picture.

\begin{figure}[t]
\begin{center}
\begin{minipage}[t]{8 cm}
\epsfig{file=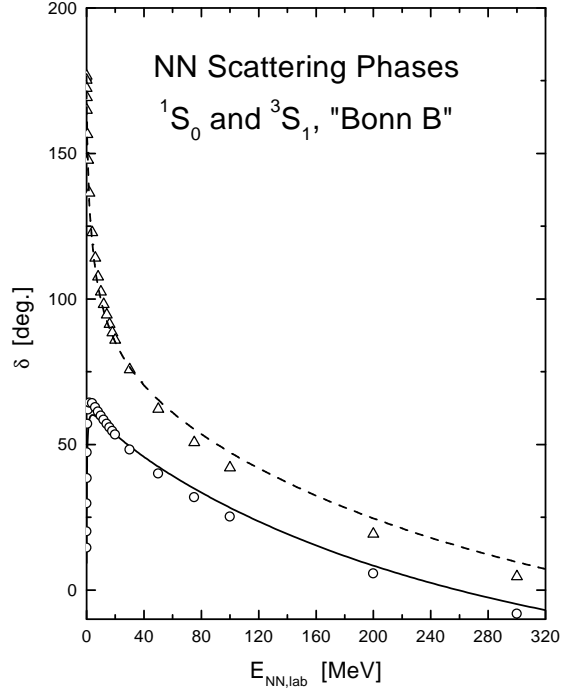,width=8cm}
\end{minipage}
\begin{minipage}[t]{16.5 cm}
\caption{Singlet-Even ($^1S_0$, full line) and Triplet-Even
($^3S_1$, dashed line) S-wave phase shifts obtained with the
D3Y-interaction in free space are compared to calculations with
the Bonn-B NN potential (circles and triangles)
\protect\cite{Machleidt}. \label{fig2}}
\end{minipage}
\end{center}
\end{figure}

A necessary condition for particle-stability is $\lambda <0$.
Then, irrespective of the value of $E$, the hole-type solutions
$\Phi_{-}$ are exponentially decaying for $r \to \infty$. For
E$\leq |\lambda|$, an exponential asymptotic behaviour is also
found for the particle-type components $\Phi_+$ and a discrete
subset of eigenvalues $E$ is obtained. For E$> |\lambda|$,
eq.(\ref{gorkov}) has to be solved with continuum wave boundary
conditions for $\Phi_{+}$ leading to single particle spectral
functions being distributed continuously in energy. Hence the
quasi-particle picture, underlying BCS theory and, to some degree,
also discretized HFB theory, is extended to a fully dynamical
continuum description.

Here, we consider only like-particle $(S=0,T=1)$ pairing. Writing
for protons ($q=p$) and neutrons ($q=n$)
$\Phi_+(r,e_+)=u_q(e_+)F_q(r,e_+)$ and
$\Phi_-(r,e_-)=v_q(e_-)G_q(r;e_-)$, respectively, where the
reduced wave functions $F_q$ and $G_q$ are normalized to unity and
$u_q,v_q$ correspond to BCS amplitudes, the pairing fields are
then defined in terms of the anomal or pairing density matrices
for protons and neutrons, respectively,
\begin{equation}
\kappa_q(r_1,r_2)=\frac{1}{2}\sum_{j\ell}{\frac{2j+1}{4\pi}\int^\infty_0{dE
u_{qj\ell}(E)v_{qj\ell}(E)F_{qj\ell}(r_1,E)}G_{qj\ell}(r_1,E)}
\end{equation}
and the proton and neutron pairing fields are
\begin{equation}
\Delta_q(r_1,r_2)=V_{SE}(r_1,r_2)\kappa_q(r_1,r_2) \quad .
\end{equation}
If $\lambda<0$, $\kappa_q(r)$ and $\Delta(r)$ are guaranteed to
decay exponentially for $r \to \infty$. The pairing interaction in
the  $particle-particle$ channel is described by $V_{SE}$. It is
still an open question whether the free space or an in-medium
singlet-even interaction should be used (see e.g.
\cite{discrete}).

Here, for the numerical calculations a local momentum
approximation is introduced by averaging in momentum space over
the non-locality and replacing $V_{SE}(r_1,r_2) \to
M_{SE}(k_F(r))\delta(r_1-r_2)$. This corresponds to including the
off-shell correlations into the strength factor $M_{SE}$ which, as
a consequence, acquires an intrinsic density dependence. It is
given by the in-medium on-shell $SE$ scattering amplitude at the
local on-shell momentum $k_F(r)$, i.e. the (two-body) S-wave
component of the $SE$ Brueckner G-matrix in the
$particle-particle$ channel. The density dependence is re-adjusted
such that for $\rho \to 0$ the free space $SE$ S-wave phase shifts
are reproduced. A comparison to the Bonn-B phase shifts
\cite{Machleidt} is shown in Fig.\ref{fig2}.

\begin{figure}[t]
\begin{center}
\begin{minipage}[t]{8 cm}
\epsfig{file=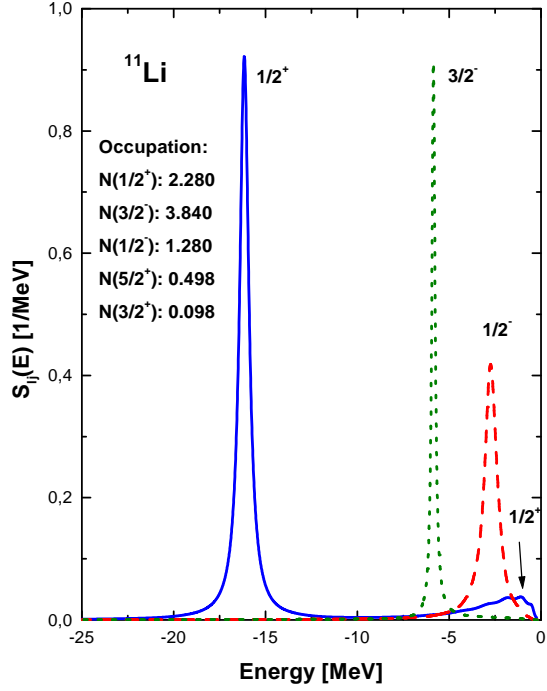,width=8cm}
\end{minipage}
\begin{minipage}[t]{16.5 cm}
\caption{Single particle spectral functions for s- and p-wave
neutron states in $^{11}$Li. The finite widths of the states is
due to continuum coupling. The partial occupation numbers
$N_{j\ell}$ are also shown, including d-wave contributions.
\label{fig3}}
\end{minipage}
\end{center}
\end{figure}

In order to understand pairing in weakly bound nuclei in more
detail it is instructive to extract from eq.(\ref{gorkov})
(non-local and energy-dependent) dynamical pairing self-energies
in the particle and hole channel, respectively,
\begin{equation}
\Sigma_\pm(1,2;\omega)=\Delta(1)^\dag g_\mp(1,2;\omega) \Delta(2)
\quad ,
\end{equation}
where $g_\pm(1,2;\omega)$ is the mean-field single particle
propagator in the complementary (hole or particle) channel. This
leads to the equivalent set of (formally) decoupled
integro-differential equations
\begin{equation} \label{decouple}
(h(1)-e_{+}+\int{d2\Sigma_+(1,2;E)})\Phi_{+} = 0 \quad ,\quad
(h(1)-e_{-}+\int{d2\Sigma_-(1,2;E)})\Phi_{-} = 0
\end{equation}
showing that the generalized pairing approach may depart strongly
from a conventional BCS description, especially when $\Sigma_\pm
\sim {\mathcal O}(e_\pm)$ becomes comparable in magnitude to the
effective single particle energies $e_\pm$. In the constant-gap
BCS approximation $\Delta(r) \to -\Delta_0$ the self-energies
become
\begin{equation}
\Sigma^{BCS}_+=-\Delta_0\frac{v_{j\ell}}{u_{j\ell}} \quad ; \quad
\Sigma^{BCS}_-=-\Delta_0\frac{u_{j\ell}}{v_{j\ell}}
\end{equation}
and eq.(\ref{decouple}) reduces to the conventional BCS eigenvalue
problem.

Returning to the full problem, strength functions for $^{11}$Li
are displayed in Fig.\ref{fig3}. The strong deviation from a pure
mean-field or BCS description is apparent by observing that
besides the expected s- and p-wave components also $d_{5/2,3/2}$
strength is lowered into the bound state region. Remarkably, the
mean-field does not support neither bound $2s$ nor $1d_{5/2,3/2}$
single particle levels and their appearance is solely due to
pairing. The states acquire a finite width because of the coupling
to the particle continuum.

\begin{figure}[t]
\begin{center}
\begin{minipage}[t]{8 cm}
\epsfig{file=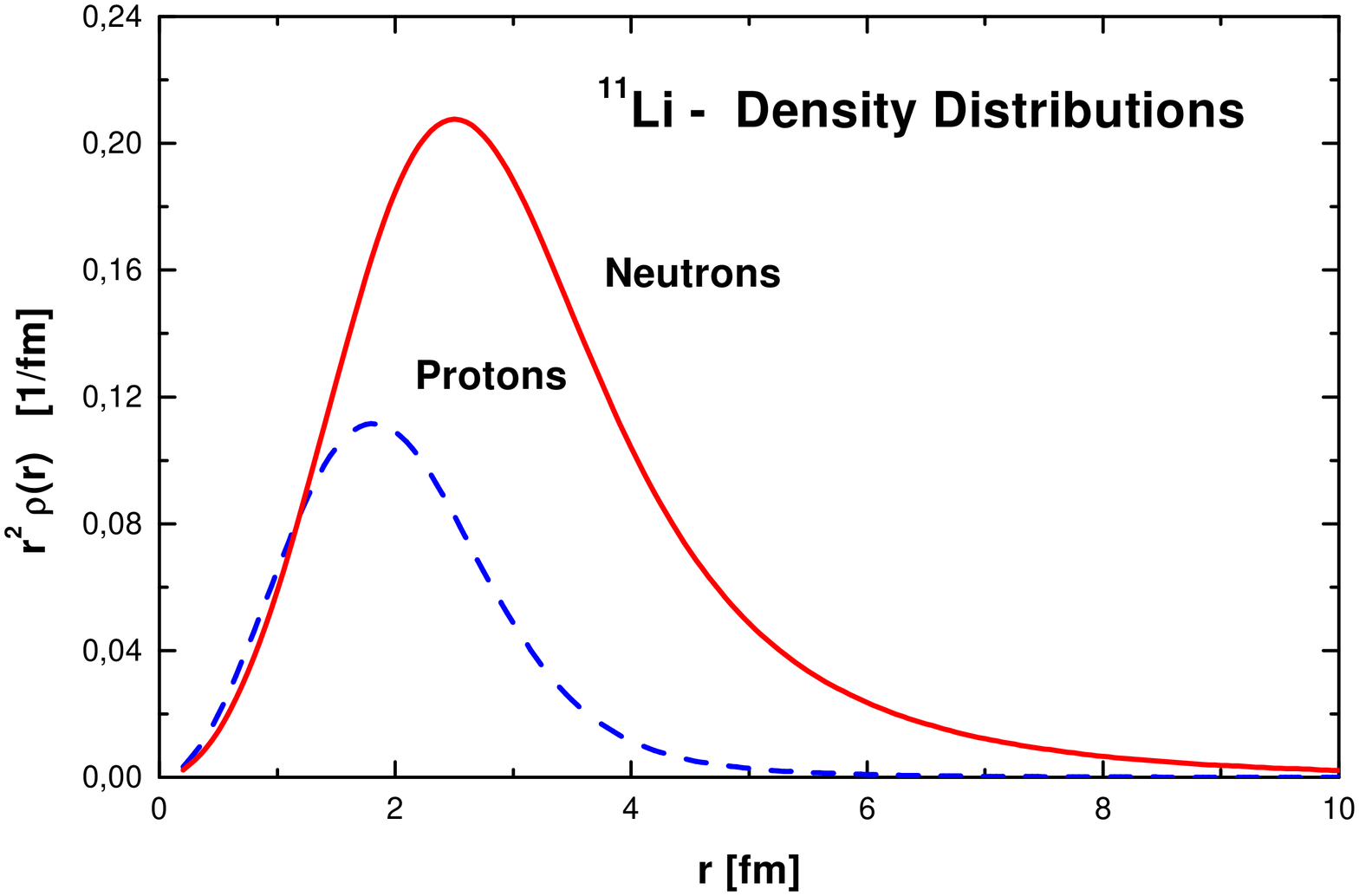,width=8cm}
\end{minipage}
\begin{minipage}[t]{16.5 cm}
\caption{Proton and neutron ground state density distributions
(weighted by $r^2$) for $^{11}$Li. The strong neutron halo
component formed by s-,p- and d-states is clearly visible.
\label{fig4}}
\end{minipage}
\end{center}
\end{figure}

Theoretically, the proton ($q=p$) and neutron ($q=n$) densities in
a systems like $^{11}$Li are defined by
\begin{equation}\label{density}
\rho_q(r)=\sum_{j\ell}{\frac{2j+1}{4\pi}\int^\lambda_{-\infty}{de_-
v^2_{qj\ell}(e_-)|G_{qj\ell}(r,e_-)|^2}} \quad ,
\end{equation}
from which the particle numbers
\begin{equation}
N_q=\sum_{j\ell}{(2j+1)\int^\lambda_{-\infty}{de_-
v^2_{qj\ell}(e_-)}}=\sum_{j\ell}{(2j+1)n_{qj\ell}} \quad ,
\end{equation}
are found. The neutron partial wave occupation numbers
$N_{j\ell}=(2j+1)n_{j\ell}$ are included in Fig.\ref{fig3}. In
stable nuclei discrete levels at $2\lambda < e_- < \lambda$ will
contribute to eq.(\ref{density}). In extreme dripline nuclei they
are missing because of the smallness of $|\lambda|$. In
Fig.\ref{fig4} the proton and neutron ground state densities are
shown, multiplied by $r^2$ in order to emphasize the differences
in shape and the neutron halo component. Applications of the
densities in high-energy elastic scattering of $^{11}$Li on a
proton target are discussed in \cite{egelh}. Measured $^{11}$Li
response functions were analyzed by QRPA calculations in
\cite{Zinser}.

Rather than solving the full continuum pairing problem as
discussed here a widely used approach is to represent the Gorkov
spectrum by a set of discretized energy levels by imposing a box
boundary condition on the solutions of eq.(\ref{gorkov})
\cite{discrete}. A comparison shows a good agreement between the
full and the discretized calculations for binding energies,
chemical potentials and ground state densities provided the box
radius is chosen large enough ($R_{box}\geq 50fm$).

\section{Dynamical Core Polarization at the Dripline}

Dynamical core polarization is seen most clearly in nuclei with a
single nucleon outside a core. Approaching the driplines the core
nucleus by itself is already far off stability containing weakly
bound single particle orbits. Under these circumstances the
surface tension and therefore the restoring forces against
external perturbations are reduced. Such "soft core" systems are,
for example, found in the neutron-rich even-mass carbon isotopes.
A good indicator is the existence of low-energy 2$^+$ states,
decreasing in energy with increasing mass. Continuum QRPA
calculations with a residual interaction derived in Landau-Fermi
liquid theory from the D3Y in-medium interaction reproduce the
systematics of 2$^+$ states rather well. This leads to the
conclusion that they are mainly of vibrational nature rather than
due to static deformation as assumed e.g. in \cite{nun}.

The calculations predict a strong increase of the quadrupole
polarizibilities in the carbon isotopes for increasing neutron
excess with a maximum around $^{16,18}$C. This behaviour is due to
the lowering of the first 2$^+$ state which in $^{12}$C is located
at $E_x=4.44$~MeV and moves down to $E_x=1.66$~MeV in $^{18}$C.
The dipole (or eletric) polarizibility, however, changes only
within 10\% over the $^{10-22}$C isotopic chain.

\begin{figure}[t]
\begin{center}
\begin{minipage}[t]{8 cm}
\epsfig{file=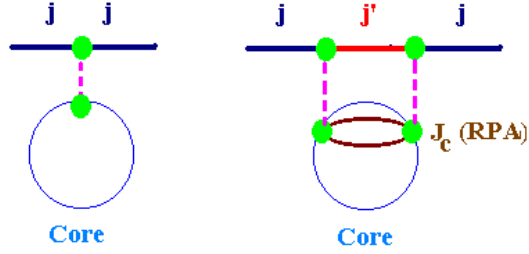,scale=0.7}
\end{minipage}
\begin{minipage}[t]{16.5 cm}
\caption{Diagramatic structure of mean-field (left) and core
polarization interactions (right) of a nucleon in single particle
state $j$. Interactions (meson exchange) are indicated by dashed
lines. Core polarization leads to intermediate states $(j'J_c)$
with a (QRPA) core excitation $J_c$ and a single particle state
$j'$. \label{fig5}}
\end{minipage}
\end{center}
\end{figure}

In a system with core polarization the valence particle (or hole)
obeys a non-static wave equation
\begin{equation}
\left ( H_{MF}+\Sigma_{pol}(\varepsilon)-\varepsilon \right
)\Psi=0
\end{equation}
including the static (HFB) mean-field Hamiltonian $H_{MF}$ and the
non-local and energy-dependent polarization self-energy
$\Sigma_{pol}$ describing the rescattering of the nucleon off the
core thereby exciting it into states of various multipolarities
and excitation energies followed by subsequent de-excitations back
to the ground state (see Fig.\ref{fig5}).

During these processes the particle can be scattered virtually
into high lying orbitals. The quantum numbers of the intermediate
$2p1h$ (or $1p2h$) configurations are only constraint by the
requirement that spin and parity much match those of the leading
particle configuration, given by the state moving with respect to
the inert core in its ground state. In nuclear matter and finite
nuclei these processes are known to give rise to a depletion of
momentum space ground state occupation probabilities
\cite{dejong96,lehr00} from the step function momentum
distribution of a Fermi gas of quasiparticles interacting only by
a static mean-field. In practice, the core excitations are
calculated in QRPA theory thus extending the static HFB picture in
a consistent way to a dynamical theory.

\begin{figure}[t]
\begin{center}
\begin{minipage}[t]{8 cm}
\epsfig{file=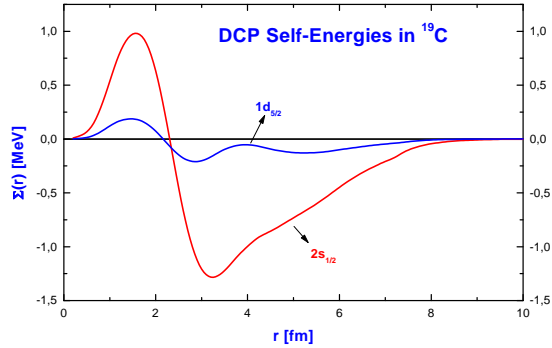,width=8cm}
\end{minipage}
\begin{minipage}[t]{16.5 cm}
\caption{Core polarization self-energies for low energy
$^{19}$C$(1/2^+)$ and $^{19}$C$(5/2^+)$ states. Only the local
parts ($r=r_1=r_2$) are shown. \label{fig6}}
\end{minipage}
\end{center}
\end{figure}

In a finite nucleus the dynamical single particle self-energies
$\Sigma_{pol}$ affect separation energies and wave functions. In
dripline nuclei the polarization self-energies are found to have a
pronounced state-dependence. In Fig.\ref{fig6} DCP self-energies
for the lowest $1/2^+$ and $5/2^+$ in $^{19}$C are shown. In the
$5/2^+$ channel the overall strength is small and a fluctuating
shape is obtained. The $1/2^+$ states, however, experience an
additional attraction from an attractive surface-type self-energy
which, in fact, provides the main source of binding for
$^{19}$C$(1/2^+,g.s.)$.

Theoretically, this amounts to use a multi-configuration ground
state containing a single particle component - reminiscent of the
static mean-field configuration - and a multitude of
configurations where the valence particle is rescattered into
other orbits by interactions with the core \cite{le,dcp}. Coupling
to the lowest 2$^+$ and 3$^-$ states only as advocated e.g. in
\cite{vib} cannot account for the complexity of the process.

\begin{table}[t]
\begin{center}
\begin{minipage}[t]{16.5 cm}
\caption{Energies, spins and ground state spectroscopic factors
from core polarization calculations. \label{tab1}}
\end{minipage}
\begin{tabular}{|c|c|c|c|}
\hline \raisebox{0pt}[13pt][7pt]{$Nucleus$}
&\raisebox{0pt}[13pt][7pt]{$j^\pi$} &
\raisebox{0pt}[13pt][7pt]{$Energy \; keV$} &\raisebox{0pt}[13pt][7pt]{$S(j^\pi,g.s.)$}\\
\hline
$^{8}B$   &$3/2^-$ & $130$ & $0.73$ \\
$^{11}Be$   &$1/2^+$ & $510$ & $0.74$ \\
$^{17}C$   &$5/2^+$ & $760$ & $0.51$ \\
s$^{19}C$   &$1/2^+$ & $263$ & $0.41$ \\
\hline
\end{tabular}
\end{center}
\end{table}

Results for energies and spectroscopic factors in the single
neutron halo nuclei $^{11}$Be and $^{19}$C and the single proton
halo nucleus $^{8}$B \cite{b8} are shown in Tab.\ref{tab1}. In
$^{11}$Be core polarization is causing the reversal of $1/2^+$ and
$1/2^-$ states by supplying an additional attractive self-energy
in the 1/2$^+$ channel. The single particle spectroscopic factor
S$_n(1/2^*,g.s.)$=0.75 is in good agreement with recent transfer
and breakup data. A more dramatic effect is found in $^{19}$C and
also $^{17}$C where the g.s. components are strongly suppressed as
seen from the small spectroscopic factors in Tab.\ref{tab1}.
Hence, the valence nucleon is no longer attached to a definite
mean-field orbital but exists in a wave packet-like state spread
over a certain range of shell model states. In other words,
mean-field dynamics have ceased to be the dominant source of
binding. Rather, the $^{17}$C and $^{19}$C results indicate a new
type of binding mechanism in dripline nuclei where shell
structures are dissolved and binding is obtained from dynamical
valence-core interactions.

\begin{figure}[t]
\begin{center}
\begin{minipage}[t]{8 cm}
\epsfig{file=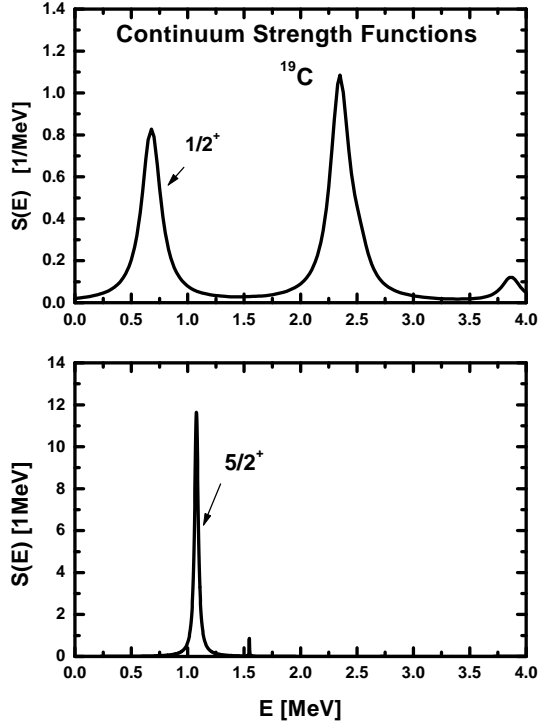,width=8cm}
\end{minipage}
\begin{minipage}[t]{16.5 cm}
\caption{Continuum strength functions for $^{19}$C$(1/2^+)$ (upper
panel) and $^{19}$C$(5/2^+)$ (lower panel). BSEC-like resonance
structures are seen, especially in $1/2^+$ channel. \label{fig7}}
\end{minipage}
\end{center}
\end{figure}

Core polarization also affects the low energy continuum region of
dripline nuclei. The admixtures of core excitations may give rise
to configurations where neither of the involved nucleons is in a
state above threshold although the total energy is well above
particle threshold. Such states are known as {\it bound states
embedded in the continuum} (BSEC). They decay by coupling to
energetically degenerate single particle continuum states. A known
case is a $3/2^+$ state in $^{13}$C which was predicted in
\cite{baur} and observed experimentally \cite{BSEC}. In $^{19}$C
BSEC-type configurations are indeed found in the s-,p- and d-wave
channels below about 6~MeV above neutron threshold. They appear as
resonances in the continuum strength functions. In Fig.\ref{fig7}
$^{19}$C$(1/2^+)$ and $^{19}$C$(5/2^+)$ DCP continuum spectral
functions are shown. The BSEC structures are most prominent in the
$1/2^+$ channel because the missing potential barrier will not
support potential resonances. In other words, observing a $1/2^+$
resonance is a direct proof for a state with a non-trivial
many-body structure. The $5/2^+$ state seen in Fig.\ref{fig7} is
part of a (much broader) conventional d-wave potential resonance
being shifted into the low energy region by core interactions.
Clearly, the existence of such close-to-threshold BSEC states will
have important consequences for breakup reactions and, especially,
for neutron capture in astrophysical processes.

The theoretical overlap wave functions have been used in eikonal
breakup calculations at relativistic energies. The FRS/GSI data
for longitudinal momentum distributions \cite{b8,c19} and removal
cross sections \cite{lola} are well described as seen from Table
\ref{tab2}.

\begin{table}[t]
\begin{center}
\begin{minipage}[t]{16.5 cm}
\caption{Theoretical one-nucleon removal cross section on a
$^{12}$C target and width of the calculated momentum distributions
obtained with core-polarized wave functions and eikonal reaction
calculations are compared to data \protect\cite{lola}. One-proton
and one-neutron removal reactions are indicated by (-1p) and
(-1n), respectively. \label{tab2}}
\end{minipage}
\begin{tabular}{|c|c|c|c|c|c|}
\hline \raisebox{0pt}[13pt][7pt] {$Isotope$} &
\raisebox{0pt}[13pt][7pt] {$Energy$} & \raisebox{0pt}[13pt][7pt]
{$\sigma_{-1N}$} & \raisebox{0pt}[13pt][7pt] {$FWHM$} &
\raisebox{0pt}[13pt][7pt] {$\sigma_{-1N}^{exp}$} &
\raisebox{0pt}[13pt][7pt] {$FWHM^{exp}$}\\
\hline
       & \raisebox{0pt}[13pt][7pt] {$MeV/nucleon$}  &
       \raisebox{0pt}[13pt][7pt] {$mb$}
       & \raisebox{0pt}[13pt][7pt] {$MeV/c$} & \raisebox{0pt}[13pt][7pt] {$mb$}
       & \raisebox{0pt}[13pt][7pt] {$MeV/c$} \\
       \hline
$^8B$ $(-1p)$ & 1440   & 104 & 75 & 98$\pm$6 & 91$\pm$5 \\
$^{10}B$ $(-1p)$ & 1450 & 17.3 & 145 & 17$\pm$2 & 165$\pm$8 \\
$^{17}C$ $(-1n)$ & 904  & 124 & 132  & 129$\pm$22 & 143$\pm$5 \\
$^{19}C$ $(-1n)$ & 910  & 192 & 69  & 233$\pm$51 & 68$\pm$3\\
\hline
\end{tabular}\\
\end{center}
\end{table}

\section{Summary}

Nuclear many-body theory was applied to strongly asymmetric
nuclear matter and the ground and excited states of dripline
nuclei. Interactions were derived from Brueckner and
Dirac-Brueckner calculations supplemented by additional density
dependences accounting for interaction contributions which are not
contained in the ladder approach. Pairing in weakly bound system
was described in a approach taking into account the coupling to
unbound continuum configurations. Dynamical core polarization in
single nucleon halo systems was investigated by HFB and QRPA
methods, also accounting for continuum effects.

The overall features of dripline nuclei are rather well described.
However, open and interesting questions remain on the structure of
interactions especially in the low-density region. Here, the
present treatment of interactions lacks full self-consistency by
the semi-empirical adjustments to variational results. Especially
three-body interactions and induced 2-body interactions from ring
diagrams in weakly bound asymmetric nuclear matter and finite
nuclei have to be investigated further. Experimentally, this could
be complemented by high-resolution measurements of spectral
functions in the bound and the continuum region.

The calculations strongly emphasize the importance of continuum
coupling. They are found to dissolve shell structures close to the
driplines and create BSEC structures in the low energy continuum.
It will be interesting to investigate core polarization and
continuum pairing for transfer and breakup reactions and capture
reactions in astrophysical scenarios.

\section*{Acknowledgments} This work is supported in part by
DFG (contract Le439/4), GSI Darmstadt and BMBF. Inspiring
collaborations with the GSI/FRS nuclear structure groups and G.
Schrieder, TU Darmstadt, are gratefully acknowledged.

\end{document}